\documentclass[preprintnumbers, prd, onecolumn, floatfix,amsmath,amsmath,
preprintnumbers,showpacs, letterpaper, superscriptaddress,nofootinbib]{revtex4}
\usepackage{graphicx,epsfig}
\usepackage{amsmath}
\usepackage {amssymb}
\usepackage{subfigure}
\usepackage{color}
\usepackage{amsfonts}
\usepackage[left=1.5cm,right=1.5cm,top=1.5cm,bottom=1.5cm]{geometry}
\usepackage[usenames,dvipsnames,svgnames]{xcolor}  

\usepackage{hyperref}   
\hypersetup{colorlinks=true,linkcolor=beamer@PRD, citecolor=beamer@PRD}
\definecolor{beamer@PRD}{RGB}{46,48,146}

\def \be  {\begin{equation}}
\def \ee  {\end{equation}}
\def \ee  {\end{equation}}
\def \bea {\begin{eqnarray}}
\def \eea {\end{eqnarray}}

\newcommand{\nn}{\nonumber}

\def\be {\begin{equation}}
\def\ee {\end{equation}}
\def\bea {\begin{eqnarray}}
\def\eea {\end{eqnarray}}
\def\bc {\begin{center}}
\def\ec {\end{center}}
\def\bfg {\begin{figure}}
\def\efg {\end{figure}}
\def\bi {\begin{itemize}}
\def\ei {\end{itemize}}
\def\nn {\nonumber}

\def\la {\label}
\def\le {\left}
\def\ri {\right}

\def\beq{\begin{equation}}
\def\eeq{\end{equation}}
\def\br{\begin{eqnarray}}
\def\er{\end{eqnarray}}
\newcommand{\eel}[1] {\label{#1}\end{equation}}
\begin{document}

\title{Planck scale effects on the stochastic gravitational wave background
generated from cosmological hadronization transition: A qualitative study }

\author {\textbf{Mohsen Khodadi}}
\email{m.khodadi@stu.umz.ac.ir}
\affiliation{Department of Physics, Faculty of Basic Sciences,\\
University of Mazandaran, P. O. Box 47416-95447, Babolsar, Iran}

\author {\textbf{Kourosh Nozari}}
\email{knozari@umz.ac.ir }
\affiliation{Department of Physics, Faculty of Basic Sciences,\\
University of Mazandaran, P. O. Box 47416-95447, Babolsar, Iran}
\affiliation{Research Institute for Astronomy and Astrophysics of Maragha
(RIAAM), P. O. Box 55134-441, Maragha, Iran}

\author{\textbf{Habib Abedi}}
\email{h.abedi@ut.ac.ir}
\affiliation{Department of Physics, University of Tehran, North Kargar Avenue, 14399-55961 Tehran, Iran}

\author{\textbf{Salvatore Capozziello}}
\email{capozziello@na.infn.it}
\affiliation{Dipartimento di Fisica ``E. Pancini``, Universit\'a di Napoli ``Federico II'',
Complesso Universitario di Monte Sant' Angelo, Edificio G, Via Cinthia, I-80126, Napoli, Italy\\
 Istituto Nazionale di Fisica Nucleare (INFN),  Sezione di Napoli,
Complesso Universitario di Monte Sant'Angelo, Edificio G, Via Cinthia, I-80126, Napoli, Italy\\
Gran Sasso Science Institute, Viale F. Crispi, 7, I-67100, L'Aquila, Italy.\\}

\begin{abstract}
We reconsider the stochastic gravitational wave background  spectrum produced
during  the  first order hadronization process, in  presence of ultraviolet  cutoffs
suggested by the  \emph{generalized uncertainty principle}  as a promising signature
towards the Planck scale physics. Unlike common perception that the dynamics of QCD phase
transition and its phenomenological consequences are highly influenced by the critical temperature,
we find that the underlying Planck scale modifications can affect the stochastic gravitational spectrum arising from the
QCD transition without a noteworthy change in the relevant critical temperature. Our investigation
shows that incorporating the natural cutoffs into MIT bag equation of state and background evolution
leads to a growth in the stochastic gravitational  power spectrum, while the relevant redshift of the QCD era, remains
unaltered. These results have double implications from the point of view of phenomenology. Firstly, it is
expected to enhance the chance of detecting the stochastic gravitational  signal created by such a transition in future observations.
Secondly, it gives a hint on the decoding from the dynamics of QCD phase transition.
\end{abstract}

\maketitle
\section{Introduction}
The detection of gravitational waves (GWs), as a physical phenomenon generated by some energetic
processes in the universe, have become increasingly important recently both for the
theoretical physicists and the observational astrophysicists. In the light of technological developments via the
construction of some sensitive detectors, the existence of gravitational radiation as a physical
reality has been confirmed for repeated occasions. From the experimental point of view, each of technologies
depending on their frequency sensitivities, have been designed for a specific purpose. Detectors in LIGO
and Virgo scientific collaborations \cite{ref:Ligo1, ref:Ligo2} have been designed for capture of high
frequency ($10-10^3$ HZ) GWs from compact binary inspiral events. For detection of sources with lower
frequency ($10^{-5}-1$ HZ) GWs signals from sources such as Supernovae, the ELISA experiment \cite{ref:Elisa} has been designed.
However, there are some setups as SKA \cite{ref:Ska1, ref:Ska2} and PTA \cite{ref:Pta} which are
trying to measure possible GWs generated with frequencies even lower than $10^{-5}$ HZ (in particular
around $10^{-9}$ HZ). This category of experiments shows that, although the strongest GWs are produced by catastrophic
events such as colliding black holes,  stellar core collapses (supernovae) and coalescing neutron
stars and so on, there is also the possibility of a random background of GWs, the so called
\emph{``stochastic gravitational-wave background ''} (SGWB) without any specific sharp frequency
component.  Furthermore, this background could be  related to extended theories of gravity \cite{RevNoi,RevVasilis} which could give rise to observable effects at cosmological level \cite{fRGW}.

Despite the fact that low-frequency SGWBs are hardly detectable empirically, the modeling of their
sources are of great theoretical interest  in the sense that might give vital information about the
very early stages of our universe \cite{ref:Kumar}. Based on the existing literatures, some sources such as
soliton stars, cosmic strings and cosmological phase transitions can be considered as SGWB generators,
see for instance \cite{ref:Source, ref:riv} for some good reviews reported recently. Concerning the cosmological sources,
 first order phase transitions  could give rise to the
 possibility of production of SGWB signals with very low frequency as designated for SKA and PTA experiments
\cite{ref:Cap, ref:riv}. According to the standard model of particle physics,
the universe, in the course of its evolution, has experienced transitional phases for several times due to spontaneous symmetry
breaking. Among these phase transitions,  one can point out to the electroweak
symmetry breaking via Higgs mechanism and the chiral symmetry breaking which results in quantum chromodynamics
(QCD) phase transition \cite{ref:Takeo, ref:Ahmad1, ref:Ahmad2}. About the nature of these transitions
and the exact critical temperature where they occurred,  there is no agreement so far.
The main goal in the study of the hadronization process, relevant to strong QCD phase transition, is
to reach the equation of state (EoS) governing two different phases, quark-gluon plasma (QGP) and hadronic gas (HG).
In relation to the undeniable role of the EoS in direct or sideway investigations of QCD, it has to be
emphasized that our knowledge of the different aspects of the system under study is highly affected by the choice of EoS
\cite{ref:Kumar, ref:Hajkarim}. Although some phenomenological models point out that  the nature of hadronization process
should be a crossover \cite{ref:Cross1, ref:Cross2, ref:Cross3}, there are other models for describing
the collective flow in heavy ion collisions which suggest EoS in a first order phase transition
with  a critical temperature around $\sim0.2$ GeV \cite{ref:QCD1, ref:QCD2, ref:QCD3}.

Given the important phenomenological role of EoS, here, we want to revisit the relevant SGWB
power spectrum arising from de-confinement to confinement first order transition in the light of
Planck scale modified EoS released in the MIT bag phenomenological model \cite{ref:Bag} where the
chemical potential is set to be zero. In particular, we utilize the proposed ``generalized uncertainty principle''
(GUP) which via the following deformation
\bea
[x_i, p_j]\hspace{-1ex} &=&\hspace{-1ex} i \hbar\hspace{-0.5ex} \left[  \delta_{ij}\hspace{-0.5ex}
- \hspace{-0.5ex} \alpha \hspace{-0.5ex}  \le( p \delta_{ij} +
\frac{p_i p_j}{p} \ri)
+ \alpha^2 \hspace{-0.5ex}
\le( p^2 \delta_{ij}  + 3 p_{i} p_{j} \ri) \hspace{-0.5ex} \ri], \label{eq:alfaa}
\label{comm01}
\eea predicts two natural ultraviolet (UV) cutoffs as a minimal physical length ($x_{min} ~
\approx ~ \alpha_0\ell_{p}$) as well as a maximum physical momentum ($p_{max} ~\approx ~ \frac{M_{p}c}
{\alpha_0}$) in contrary to the standard quantum commutator relation governing the micro-scale particles i.e.
Heisenberg uncertainty principle (HUP) \cite{ref:GUP1, ref:GUP2, gae}. It is clear that the above deformed
commutator relation is different  from HUP via Planck scale characteristic parameter $\alpha=\frac{
\alpha_0}{M_pc}=\frac{\alpha_0\ell_p}{\hbar}$,  where $\alpha_0, M_p,~l_p$ denote a dimensionless
constant, Planck mass and Planck length, respectively. By setting $c=1$ (speed of light),
one finds $\alpha=\frac{\alpha_0}{E_p}$ which $E_p$ is Planck energy $\approx10^{19}$ GeV. Due to the importance
of UV cutoffs in the vicinity of Planck scale,  quantum gravity (QG) models impose the
order of magnitude of unity and subsequently $10^{-19}$ GeV$^{-1}$ for $\alpha_0$ and $\alpha$,
respectively. By involving QG considerations in a common physical framework, so far researchers were not able
to derive upper bound better than $\alpha_0<10^{10}$ (equivalent to $\alpha<10^{-9}$
GeV$^{-1}$), \cite{ref:DasPRL}. See also \cite{ref:Upper1,ref:Upper2,ref:Upper3,ref:Upper4,ref:Upper5,ref:Upper6}
for other upper bounds obtained further in different contexts. However, unlike the prevailing
notion that, in the context of today's technology, it is impossible to probe the Planck scale physics
with ideal resolution, in Refs. \cite{ref:Nature, ref:Sci, ref:NPB}, by employing an
opto-mechanical setup, surprisingly it has been shown that this ideal is achievable via a simple
table-top experiment. This is a success for QG in the sense that $\alpha_0=1$ ( or $\alpha=10^{-19}$ Ge$V^{-1}$)
is no longer merely a theoretical ideal out of reach of today's technology. We note also
that in Ref. \cite{ref:natur2013}, the authors by considering the Planck scale physics into the gravitational bar detectors, have been
succeeded in extracting a solid constraint for possible Planck-scale corrections on the ground-level energy of an oscillator.
\\

This idea that GUP can affect the dynamics of QCD phase transition comes back to the fact
that GUP, by modifying the fundamental commutator bracket between position and momentum
operators, leads to some modifications in the Hamiltonian of physical systems \cite{ref:DasPRL}.
Therefore, GUP, by imposing
its contribution within the thermodynamical quantities relevant to QGP and hadron
phases, is expected to affect  the hadronization process. Besides, studies carried out
so far, confirm the point that the natural cutoff approaches make remarkable contributions
in cosmological and astrophysical systems, for instance we can mention  the Refs.
\cite{ref:CQG, ref:PLB, ref:JCAP, ref:PLB2016}. In this paper, the QGP phase includes two
massless quark flavors with zero chemical potential which at a HG-QGP phase
thermodynamic equilibrium is affected by the underlying GUP modifications.
As a result, the relevant thermodynamic quantities are derived in the presence of
the above mentioned natural UV cutoffs. Finally, by means of the effective degrees of
freedom and MIT bag pressure the confinement hadronic phase shall be distinct from
the de-confined QGP. So, as discussed in section \ref{sec:Thermo}, for each phase we
deal with the GUP-modified EoS(s), separately. We emphasize that, for the
present paper, our main focus is on the relevant EoS of QGP phase, because we are interest
in the SGWB from phase transition above the critical temperature. Then, by using the GUP-modified
QCD EoS derived in section \ref{sec:Thermo}, we calculate the modified general expression of
SGWB power spectrum in section \ref{sec:SGWB}. The discussion is expanded to section
\ref{sec:Sources} by introducing some QCD sources which during first order transition production of SGWB can
be involved. Finally, we provide a summary of the present work along with
conclusions in section \ref{sec:Result}.

\section{quark-gluon EOS in GUP Model including minimal length and maximal momentum invariant cutoffs}
\label{sec:Thermo}
\begin{figure}
\begin{center}
 \epsfig{figure=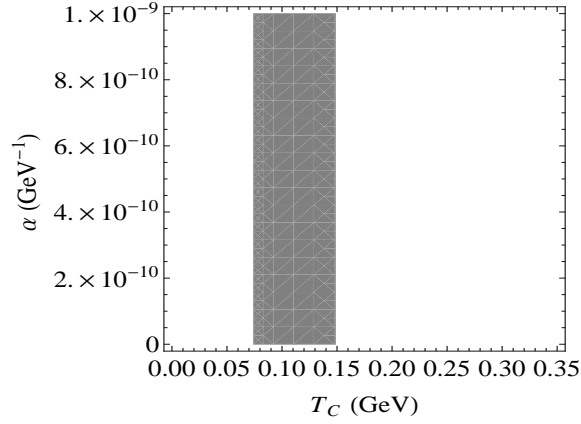,width=3in,height=2.2in,angle=0}
\end{center}
\caption{\emph{A diagram of $(T_c,\alpha)$ parameter space. $T_c$ is constrained to the interval $0.07-0.15$ GeV}}
\label{Bag}
\end{figure}

Let us introduce UV cutoffs modified thermodynamics quantities such as pressure
$P$, energy density $\rho$ and entropy density $s$ for QGP and Hadronic Gas HG phases.
Concerning the GUP model at hand, for a massless particle as Pion in HG state, the
standard dispersion
relation $ E^{2}(k)=k^{2}$ modifies as \cite{ref:Ali, ref:Das, ref:Vagenas, ref:Elmash}
\begin{equation}\label{e2-0}
E^{2}(k)=k^{2}(1-2\alpha {k}),
\end{equation} for energies close to Planck scale\footnote{Of course, in the presence of only the minimum length, the correction term in Eq. (\ref{e2-0}), appears with positive sign \cite{ref:ahep2015, ref:jmp2013}. It is interesting to know that the GUP arising from (\ref{comm01}) is capable to provide necessary mechanism to produce more massive particles, see \cite{ref:plb2013} for more details.}. Given that the GUP affects  the measure
of integral, for a large volume of all possible Pion gas the standard partition function
extends to
\begin{eqnarray}\label{e2-1}
\ln z_{B}&=& -\dfrac{V g_\pi}{2 \pi^{2}} \int_{0}^{\infty} k^{2} \frac{\ln \le[1-\exp\le(-\frac{E(k)}{T}\ri)\ri]}{(1-\alpha k)^{4}} ~dk\nonumber \\
         &=& -\dfrac{V g_\pi}{2 \pi^{2}} \int_{0}^{\infty}  k^{2} \frac{\ln \le[1-\exp\le(-\dfrac{k}{T}(1-2 \alpha k)^{1/2}\ri)\ri]}{(1-\alpha k)^{4}} ~dk,
 \end{eqnarray} since
 \begin{equation}\label{e2-2}
\sum_{k}\rightarrow{\dfrac{V}{(2\pi)^{3}}\int_{0}^{\infty}d^{3}k}\rightarrow{\dfrac{V}{2\pi^{2}}
\int_{0}^{\infty}\dfrac{k^{2}dk}{(1-\alpha{k})^{4}}}~.
\end{equation}
In Eq. (\ref{e2-1}) the factor of $g_\pi$ refers to the number of degrees of freedom in the final HG state which is
full of Pion gas. Choosing the change of variables as $x=\dfrac{k}{T}\sqrt{1-2 \alpha k}$ and by neglecting ${\mathcal O(\alpha^2)}$ terms, after some manipulation the partition function (\ref{e2-1}) reads as
\begin{equation}\label{e2-3}
\ln z_{B}=\dfrac{V g_\pi}{2\pi^{2}}\int_{0}^{\infty} \frac{1}{3} x^{3}T^{3} \dfrac{(1+\alpha xT)^{3}}{(1-\alpha xT)^{3}}\dfrac{dx}{e^{x}-1}~.
\end{equation}
By a Taylor series expansion of $\dfrac {(1+\alpha xT)^{3}}{(1-\alpha xT)^{3}}$ around zero
in above integral, we come to the final integral form of the partition function as follows
\begin{equation}\label{e2-4}
\ln z_{B}=\dfrac{V g_\pi}{6\pi^{2}}\le[\int_{0}^{\infty} T^{3} \dfrac{ x^{3} dx}{e^{x}-1}+ \int_{0}^{\infty} 6 \alpha T^{4} \dfrac{ x^{4} dx}{e^{x}-1}\ri]~.
\end{equation}
Now by using the relation between partition function and grand canonical potential, $\ln z_{B}=-\Omega/T$, one gets
\begin{eqnarray}\label{e2-4}
\dfrac{\Omega}{V} &=& -\dfrac{g_\pi}{6\pi^{2}} T^{4} [\Gamma{(5)}I(0)_{5}^{(-)}] - \alpha \dfrac{g}{\pi^{2}}
T^{5} [\Gamma{(6)}I(0)_{6}^{(-)}] \la{omega},
\end{eqnarray} which can be represented for the pressure as $P=-\frac{\Omega}{V}$.
Here, $\Gamma{(n)}=(n-1)!$ is the gamma function and $I(0)_{n}^{(\pm)}$ denote the Bose and Fermi integrals defined as
\begin{equation}
I(y)_{n+1}^{(\pm)}=\dfrac{1}{\Gamma{(n+1)}}\int_{0}^{\infty} \dfrac {x^n}{(x^{2}+y^{2})^{1/2}}\dfrac{1}
{\exp{\left[(x^{2}+y^{2})^{1/2}\right]}\pm{1}} \la{IY} dx~.
\end{equation}
Given the fact that by fixing $ y=0 $, then $I(0)_{n+1}^{(\pm)}=\dfrac{1}{n}\zeta{(n)}a_{n}^{(\pm)} \la{I0}$
with Riemann zeta function, $ \zeta{(n)}=\sum_{j=1}^{\infty} j^{(-n)} $ which is valid for integer
${(n\geq{2})}$, finally in the hadronic phase we obtain the following UV cutoffs modified expression
\be \label{eq:ppG}
P_{HG} = g_\pi\dfrac{\pi^{2}}{90}T^{4} + \frac{24g_\pi\alpha\zeta{(5)}}{\pi^{2}}T^{5}~,
\ee for the  pressure of a massless ideal HG. Now, using trace anomaly relation
$\frac{\rho - 3p}{T^4} = T\,\frac{d}{dT}\bigg(\frac{p}{T^4}\bigg)$ and also thermodynamic
formula $s=\frac{dP}{dT}$ \cite{ref:Cheng}, the relevant energy density and entropy density
can be obtained as
\begin{eqnarray}
\rho_{HG} &=& 3 g_\pi\dfrac{\pi^{2}}{90}T^{4} + \frac{72g_\pi\alpha\zeta{(5)}}{\pi^{2}}T^{5}, \label{eq:eeG}\\
s_{HG} &=& 4 g_\pi \dfrac{\pi^{2}}{90}T^{3} + \frac{120g_\pi\alpha\zeta{(5)}}{\pi^{2}}T^{4}. \label{eq:ssG}
\end{eqnarray} respectively. The above three thermodynamic equations are valid for the HG state
and also are extendable to the QGP state by regarding the bag constant $B$ and converting the HG
number of degree of freedom $g_\pi$ to its counterpart in QGP phase, $g_{QGP}$. As a result, the counterpart
of the above thermodynamic quantities in QGP phase read as
\begin{eqnarray}
P_{QGP}&=&g_{QGP}\dfrac{\pi^{2}}{90}T^{4}\le(1-\dfrac{g_{QGP}-g_{\pi}}{g_{QGP}}\tau^{4}\ri)
+\frac{24 g_{QGP}\zeta{(5)}}{\pi^{2}}\alpha T^5\le(1-\dfrac{g_{QGP}-g_{\pi}}{g_{QGP}}\tau^5\ri), \label{eq:Pp}\\
\rho_{QGP} &=&  g_{QGP}\dfrac{\pi^{2}}{90}T^{4}\le(3+\dfrac{g_{QGP}-g_{\pi}}{g_{QGP}}\tau^{4}\ri) +
\frac{24 g_{QGP}\zeta{(5)}}{\pi^{2}}\alpha T^5\le(3+\dfrac{g_{QGP}-g_{\pi}}{g_{QGP}}\tau^5\ri).\label{eq:Ee}
\end{eqnarray} where $\tau=\frac{T_c}{T}$ with $T_c$ as critical temperature and also
\begin{equation}\label{B}
B=\dfrac{(g_{QGP}-g_{\pi})\pi^{2}}{90}T_{c}^{4}+24\zeta(5)\alpha(g_{QGP}-g_{\pi}) T_{c}^{5}~.
\end{equation}
It is important to know that by computing the entropy density also one
can find the effect of modified dispersion relation (\ref{e2-0}) on the bulk to shear viscosity ratio
as an extractable quantity in relativistic heavy ion collisions
\cite{ref:Vagenas2018}. Note that in a first order phase transition, some energy will be released in the medium. Therefore, the bag constant
indeed plays the role of laten heat released during transition and also enters the pressure of the de-confined
phase which its value can be calculated via Gibbs phase equilibrium condition $P_{HG}(T_{c})=P_{QGP}(T_{c})$ i.e.
the existence of a critical temperature $T_c$ in which the pressures in the two phases are equal
\cite{ref:Gibbs}. The phenomenological fits of light hadron properties as well as low energy hadron
spectroscopy address an explicit constraint on the bag constant as $B\in(10^{-4}-16\times10^{-4})$ GeV
\cite{ref:Lee, ref:Harko, ref:Heydari-Fard, ref:Mohsen1, ref:Mohsen2}. Now, using constraints governed on
bag constant $B$ as well as the Planck scale characteristic parameter $\alpha$ (as previously discussed)
we will be able to determine the allowed range of the critical temperature $T_c$ in the presence of QG effects.
We find that $T_c$ can be one of possible values in interval $0.07-0.15$ GeV, as can be seen in Fig. \ref{Bag}.
Here, values $\geq 0.1$ GeV are admissible since according to standard literatures concerning the strong
interaction, it is expected that the hadronization process should be occured at $T_c\sim 0.1-0.2$ GeV. Now
by setting $B=16\times10^{-4}$ GeV into the above equation one gets $T_c\sim 0.15$ GeV \footnote{Among the five possible
solutions that are expected from Eq. (\ref{B}), four solutions are rejected because of negative and imaginary character.}
which in comparison with the original MIT bag critical temperature $T_{c}=\bigg(\frac{90B}{\pi^2(g_{QGP}-g_{\pi})}
\bigg)^{1/4}$ \cite{ref:Heydari-Fard, ref:Mohsen1, ref:Mohsen2} no changes are observed. Therefore, it can be said that
taking QG considerations via incorporation of some UV natural cutoffs (in particular as minimal length and maximal momentum)
in nature, has no effect on the relevant critical temperature of hadronization process in early universe.
It is clear that by ignoring the UV cutoffs modifications i.e. by setting $\lim _{\alpha\rightarrow0}$,
then Eqs. (\ref{eq:Pp}) and (\ref{eq:Ee}) recover their standard forms
\begin{eqnarray}
P_{QGP}&=&g_{QGP}\dfrac{\pi^{2}}{90}T^{4}-B \\
\rho_{QGP} &=&  g_{QGP}\dfrac{\pi^{2}}{30}T^{4} +B ,\\
\end{eqnarray} in MIT bag model which $B=\frac{(g_{QGP}-g_{\pi})\pi^2}{90}T_{c}^4$.
It should be noted that both phases involved in the hadronization transition in early universe
contain effectively massless electro-weak (EW) particles which should be taken in number of degrees
of freedom. In de-confined QGP state it includes quark ($g_q$), gluon ($g_g$) and massless EW particles
($g_{EW}$) contributions which with assumption of two flavors, three colors and fixing QCD coupling constant at
$\alpha_S=0.5-0.6$, we will have $g_{QGP}=35\pm2$ \cite{ref:book}. In the final HG phase, aside from three
light bosons (Pions, $\pi^{\pm}$ and ($\pi^0$)), the presence of heavier hadrons may contributes at $<T_c$
which regardless of them one finds for the hadronic degrees of freedom $g_{\pi}\sim3$.

\section{SGW spectrum modified with GUP-characteristic parameter $\alpha$}
\label{sec:SGWB}
In this section we discuss the SGW spectrum generated from the epoch of cosmological QCD phase
transition influenced by the QG effects towards today epoch. Indeed, we investigate the effect of Planck
scale physics on the estimation of the current observable SGWB. To describe the evolution of the universe
it is useful to track a conserved quantity as entropy which in cosmology is more informative than energy.
The entropy of the universe is dominated by the entropy of the photon bath since there are far more photons
than baryons in the universe. Any entropy production from non-equilibrium processes is therefore completely
insignificant relative to the total entropy. With a good approximation one can treat the expansion of the
universe as adiabatic ($\dot s/s=0\,$), so that the total entropy stays constant even beyond equilibrium.
Using thermodynamic relations presented in the previous section, by regarding UV cutoffs admitted by the
GUP model at hand, then the relevant entropy density reads as
\be \label{eq:Ss}
s = 4 \eta_1T^3+5\eta_2\alpha T^4
\ee in which $\eta_1=g_{s}\dfrac{\pi^{2}}{90}~,~~\eta_2=\frac{24 g_{s}\zeta{(5)}}{\pi^{2}}$
and $g_s$ is the effective number of degrees of freedom in entropy calculation.
Thereupon, using the adiabatic condition $\dot s/s=0\,$, we can
obtain the following expression for time variation of temperature
\be \label{Tdot}
\frac{dT}{dt} = -H T\,\left(\frac{\frac{12\eta_1}{T}+15\eta_2\alpha}{\frac{12\eta_1}{T}\big(1+ \frac{T}
{3g_{s}}\frac{dg_{s}}{dT}\big)
+20\eta_2\alpha\big(1+ \frac{T}{4g_{s}}\frac{dg_{s}}{dT}\big)}\right) \,,
\ee
which in the limit $\alpha\rightarrow0$, one recovers the the standard form as
\be \label{Tdot*}
\frac{dT}{dt} = -H T\,\left(1+ \frac{T}{3g_{s}}\frac{dg_{s}}{dT} \right)^{-1}\,.
\ee
By integrating Eq. (\ref{Tdot}) we arrive at
\be\label{eq:a_*}
\frac{a_*}{a_0} = \exp \bigg[\int_{T_*}^{T_0} \frac{1}{T}\left(\frac{\frac{12\eta_1}{T}
+15\eta_2\alpha}{\frac{12\eta_1}{T}\big(1+ \frac{T}{3g_{s}}\frac{dg_{s}}{dT}\big)+20\eta_2\alpha\big(1+ \frac{T}{4g_{s}}\frac{dg_{s}}{dT}\big)}\right)dT\bigg]\,,
\ee
which by taking the Boltzmann equation\footnote{The reason for using the Boltzmann equation comes back
to this fact that GWs are essentially decoupled from the rest of the universe dynamics.} as $\frac{d}{dt}(\rho_{gw}\, a^4)
= 0\, ,$ then the energy density of the GWs at the today epoch is given as
\be\label{eq:rho_gw}
\rho_{\rm gw}(T_0) = \rho_{\rm gw}(T_*)\,\exp \bigg[\int_{T_*}^{T_0} \frac{4}{T}\left(\frac{\frac{12\eta_1}
{T}+15\eta_2\alpha}{\frac{12\eta_1}{T}\big(1+ \frac{T}{3g_{s}}\frac{dg_{s}}{dT}\big)
+20\eta_2\alpha\big(1+ \frac{T}{4g_{s}}\frac{dg_{s}}{dT}\big)}\right)dT\bigg]~.
\ee
Here and also in what follows, the signs `` * '' and `` 0 '' refer to the quantities at the epochs of phase transition
and today respectively. Now by definition of the density parameter of the GWs at today and phase transition epoch
as $\Omega_{\rm gw} = \frac{\rho_{\rm gw}(T_0)}{\rho_{\rm cr}(T_0)}$ and $\Omega_{\rm gw*} = \frac{\rho_{\rm gw}(T_*)}
{\rho_{\rm cr}(T_*)}$ respectively, we have
\be\label{eq:gw_den}
\Omega_{\rm gw}=\Omega_{\rm gw*}\frac{\rho_{\rm cr}(T_*)}{\rho_{\rm cr}(T_0)}
\exp \bigg[\int_{T_*}^{T_0} \frac{4}{T}\left(\frac{\frac{12\eta_1}{T}+15\eta_2\alpha}
{\frac{12\eta_1}{T}\big(1+ \frac{T}{3g_{s}}\frac{dg_{s}}{dT}\big)
+20\eta_2\alpha\big(1+ \frac{T}{4g_{s}}\frac{dg_{s}}{dT}\big)}\right)dT\bigg]\,.
\ee
Using Eq. (\ref{Tdot}), then the energy conservation equation $\dot{\rho}+3H\rho(1+w)=0$
can be rewritten as
\be \label{eq:EC}
\frac{d\rho}{\rho} = \frac{3}{T}\,\left(1+ w\right)
  \left(\frac{\frac{12\eta_1}{T}+15\eta_2\alpha}{\frac{12\eta_1}{T}\big(1+ \frac{T}
{3g_{s}}\frac{dg_{s}}{dT}\big)
+20\eta_2\alpha\big(1+ \frac{T}{4g_{s}}\frac{dg_{s}}{dT}\big)}\right)\, dT\, ,
   \ee
where by integrating it between two intervals, radiation dominated epoch in some early
time with relevant quantities $\rho (T_r)$ and $T_r$ until phase transition epoch, one comes
to the following expression
\be \label{eq:energy}
\rho_{cr}(T_*) = \rho_{r}(T_r)\, \exp\bigg[\int_{T_r}^{T_{*}}
  \frac{3}{T}\,(1+ w )
  \left(\frac{\frac{12\eta_1}{T}+15\eta_2\alpha}{\frac{12\eta_1}{T}\big(1+ \frac{T}
{3g_{s}}\frac{dg_{s}}{dT}\big)
+20\eta_2\alpha\big(1+ \frac{T}{4g_{s}}\frac{dg_{s}}{dT}\big)}\right)\, dT\bigg]\,,
\ee
for the critical energy density at the time of phase transition. Now by introducing
the radiation density parameter $\Omega_{\rm r0} = \frac{\rho_{\rm r}(T_0)}{\rho_{\rm cr}(T_0)}$ which can be interpreted as
fractional energy density of radiation in today epoch with observational value of $\simeq 8.5\times10^{-5}$ along
with the above equation we have
\be \label{eq:HstH0}
\frac{\rho_{\rm cr }(T_*)}{\rho_{\rm cr}(T_0)}=\Omega_{r0} \bigg(\frac{a_0}{a_r}\bigg)^4
\exp\bigg[\int_{T_{r}}^{T_{*}} \frac{3}{T}\,(1+ w )
  \left(\frac{\frac{12\eta_1}{T}+15\eta_2\alpha}{\frac{12\eta_1}{T}\big(1+ \frac{T}
{3g_{s}}\frac{dg_{s}}{dT}\big)
+20\eta_2\alpha\big(1+ \frac{T}{4g_{s}}\frac{dg_{s}}{dT}\big)}\right)\, dT\bigg]\, ,
\ee where finally by inserting into (\ref{eq:gw_den}), the GW spectrum observed today
in the presence of Planck scale corrections is acquired as
\begin{align} \label{eq:gw0}
\Omega_{\rm gw} = \Omega_{r0} \Omega_{\rm gw*} & \exp\bigg[\int_{T_*}^{T_0}
 \frac{4}{T}\left(\frac{\frac{12\eta_1}{T}+15\eta_2\alpha}
{\frac{12\eta_1}{T}\big(1+ \frac{T}{3g_{s}}\frac{dg_{s}}{dT}\big)
+20\eta_2\alpha\big(1+ \frac{T}{4g_{s}}\frac{dg_{s}}{dT}\big)}\right)dT\bigg] \nn \\
  &\times \exp \bigg[\int_{T_{r}}^{T_{*}}
   \frac{3}{T}\,(1+ w )
  \left(\frac{\frac{12\eta_1}{T}+15\eta_3\alpha}{\frac{12\eta_1}{T}\big(1+ \frac{T}
{3g_{s}}\frac{dg_{s}}{dT}\big)
+20\eta_2\alpha\big(1+ \frac{T}{4g_{s}}\frac{dg_{s}}{dT}\big)}\right)\, dT\bigg]\,.
  \end{align}

  \begin{figure}
\begin{tabular}{c}\hspace{-1cm}\epsfig{figure=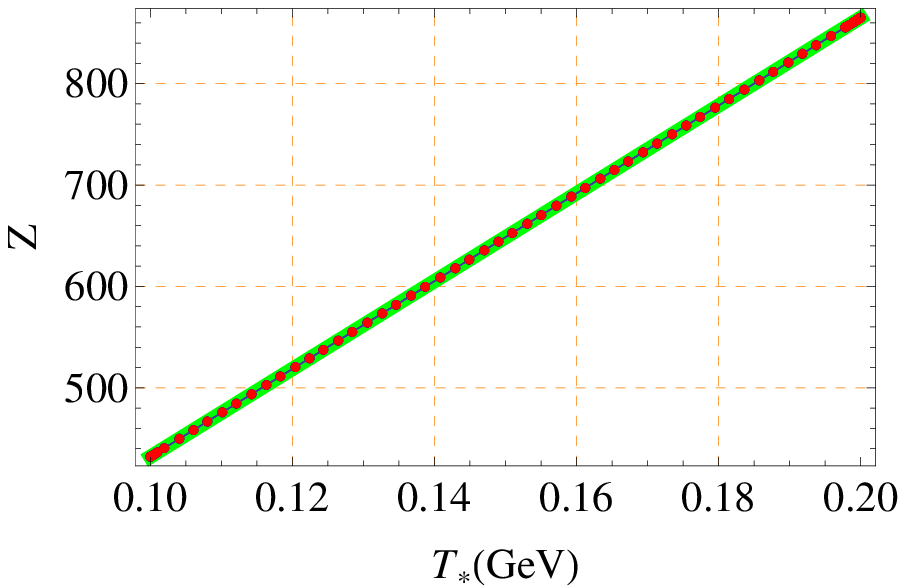, width=2.2in,height=2in,angle=0}
\hspace{0.5cm} \epsfig{figure=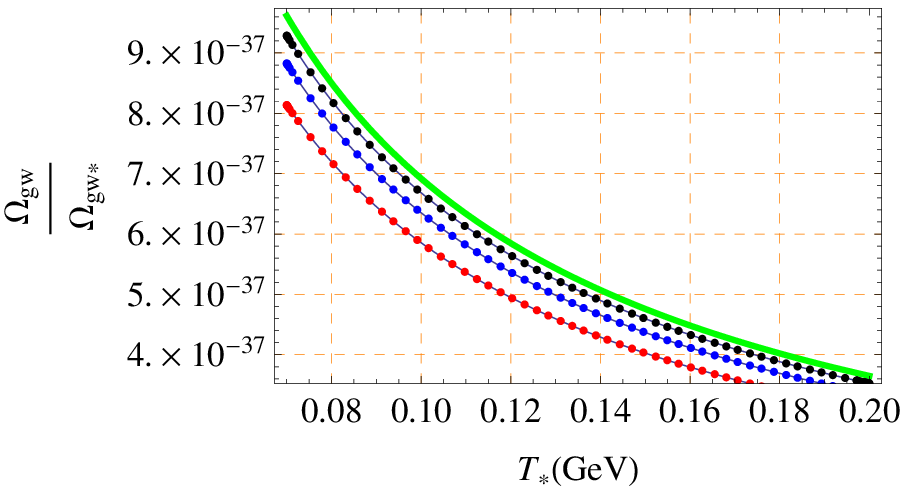, width=2.7in,height=2in,angle=0}
\hspace{0.3cm} \epsfig{figure=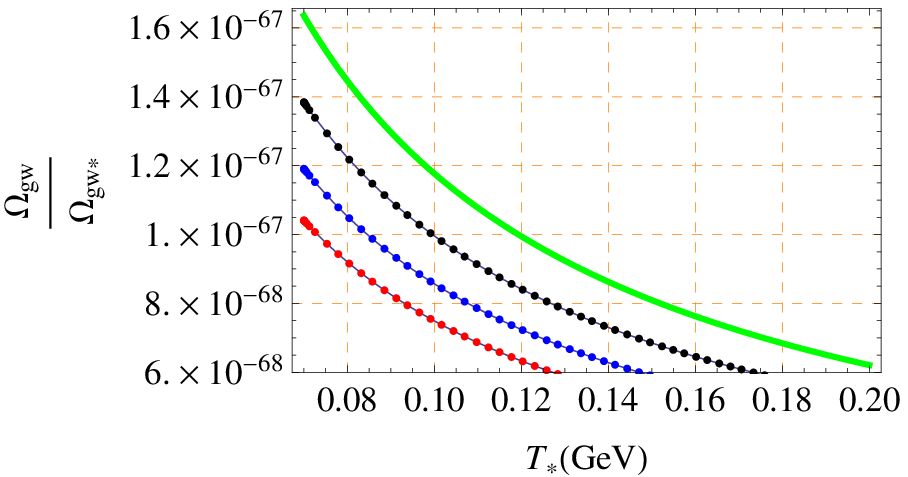, width=2.7in,height=2in,angle=0}
\end{tabular}
\caption{\emph{Left panel shows the redshift $z$ of the QCD era SGW as a function of the critical
temperature $T_c$ (equal to $T_{*}$). The middle as well as the right panels show the ratio of SGWB spectrum detected today
relative to that in the time of QCD transition as function of critical temperature $T_c$. In all three plots
the solid line with green color refers to the original bag model while the meshed lines refer to case
parametrized in the presence of UV cutoffs with different Planck scale characteristic parameter $\alpha$. }}
\label{wz}
\end{figure}

Now by using the relation between the scale factor and redshift as $\frac{a_0}{a_*}=1+z
=\frac{\nu_*}{\nu_0}$, the peak frequency of the gravitational wave red shifted to current epoch is
given by
\begin{align} \label{eq:gw-spec}
 \frac{\nu_{\rm 0(peak)}}{\nu_*} =& \bigg(\frac{a_*}{a_0}\bigg)=
 \exp\bigg[\int_{T_*}^{T_0} \frac{dT}{T}\left(\frac{\frac{12\eta_1}{T}
+15\eta_2\alpha}{\frac{12\eta_1}{T}\big(1+ \frac{T}{3g_{s}}\frac{dg_{s}}
{dT}\big)+20\eta_2\alpha\big(1+ \frac{T}{4g_{s}}\frac{dg_{s}}{dT}\big)}\right)\bigg] \nn \\
&=\frac{f_1(T_0)~g_{s}(T_0)^{\frac{f_1(T_0)}{3T_0}}~f_2(T_0)~g_{s}(T_0)^{5\alpha\eta_2f_1(T_0)}}
{f_1(T_*)~g_{s}(T_*)^{\frac{f_1(T_*)}{3T_*}}~f_2(T_*)~g_{s}(T_*)^{5\alpha\eta_2f_1(T_*)}}
\end{align} where
\be \label{eq:0}
f_1(T)=\frac{T}{12\eta_1}\big(1-\frac{5\alpha\eta_2}{4\eta_1}T)\mid_{T_0,T_*}~~~,
~~~f_2(T)=(12\eta_1)^{4/3}\big(1-\frac{20\alpha\eta_2}{9\eta_1}T)\mid_{T_0,T_*}\,.
\ee
One can show that by discarding the $\alpha$ terms, the standard form of above
equation can be recovered
\be \label{eq:1}
\frac{\nu_{\rm 0(peak)}}{\nu_*} = \frac{T_0}{T_*}.\bigg(\frac{g_{s}(T_0)}{g_{s}(T_*)}\bigg)
^{\frac{1}{36\eta_1}}\, .
\ee
In order to see the effect of UV cutoffs on the behavior of the relevant redshift $z$ of GW
spectrum, in Fig. \ref{wz} (left panel) using Eqs. (\ref{eq:gw-spec}) and (\ref{eq:1}) we
have plotted $z$ versus $T_*$ where we have used the numerical values as $ T_0=0.2348\times10^{-3}
~GeV,~g_{s}(T_*) \in [33-37],~g_{s}(T_0)=3.4$ with $\alpha=0$ (green line), $\alpha=10^{-19}~ GeV^{-1}$
(black meshed line), $\alpha=10^{-18}~ GeV^{-1}$ (blue meshed line) $\alpha=10^{-17}~ GeV^{-1}$
(red meshed line). This figure shows that incorporating the Planck scale characteristic parameter
$\alpha$ in MIT bag EOS, does not cause a significant change on redshift of the QCD era SGW.
This is reasonable in the sense that in the previous section it was shown that the relevant critical
temperature of QCD transition, has not been affected due to the addition of $\alpha$-terms
into EOS. Namely, corrections generated by the underlying GUP model does not move the
location (red shift) of phase transition relative to today observer. In what follows,
using Eq. (\ref{eq:a_*}) we reexpress Eq. (\ref{eq:gw0}) as
\begin{align} \label{eq:gw1}
\frac{\Omega_{\rm gw}}{\Omega_{\rm gw*}}=\Omega_{r0} \big(\frac{a_*}{a_0}\big)^4
\big(\frac{a_r}{a_*}\big)^{3w+3}~;
\end{align} where
\be
\big(\frac{a_r}{a_*}\big)^{3w+3}= \bigg(\frac{f_1(T_*)}{f_1(T_r)}\bigg)^{3w+3}\times
\bigg(\frac{g_{s}(T_*)^{\frac{f_1(T_*)(3w+3)}{3T_*}}}{g_{s}(T_r)^{\frac{f_1(T_r)(3w+3)}{3T_r}}}\bigg)\times
\bigg(\frac{f_2(T_*)}{f_2(T_r)}\bigg)^{3w+3}\times\bigg(\frac{g_{s}(T_*)^{5(3w+3)\alpha\eta_2f_1(T_*)}}{g_{s}(T_r)
^{5(3+3w)\alpha\eta_2f_1(T_r)}}
\bigg)~.
\ee
By focusing on the form of the above equation written in terms of critical temperature
\footnote{Note that in this paper we study the SGWBs
generated during the first order transition from QGP to HG phase. Since this event can be addressed by
the critical temperature, so $T_*=T_c$.}, we notice a
tangible change in the fractional energy density of the SGWB due to application of the UV cutoffs'
modifications to the MIT bag EOS, see Fig.\ref{wz} (middle and right panels). As is clear, the addition
of Planck scale characteristic parameter
$\alpha$ into the MIT bag EOS, leads to a reduction in SGWB signal arising from QCD phase transition. Interestingly,
we found that the deviations caused by UV cutoffs for values very close to $\alpha=10^{-19}$ GeV$^{-1}$,
are highly dependent on the value specified for $T_r$. The middle panel of Fig. \ref{wz} is depicted with fixed values $T_r=10^8$ GeV,
$ T_0=0.2348\times10^{-3} ~GeV,~g_{s}(T_r)=106,~g_{s}(T_*) \in [33-37],~g_{s}(T_0)=3.4$ with $\alpha=2\times10^{-10},
~5\times10^{-10}~, 10\times10^{-10}$ GeV$^{-1}$ from black meshed towards red meshed line respectively, while in the right panel we have set $T_r=10^{16}$ GeV
with $\alpha=1\times10^{-19},~2\times10^{-10}~, 5\times10^{-9}$ GeV $^{-1}$ with the same other numerical values.
\begin{figure}
\begin{center}
 \epsfig{figure=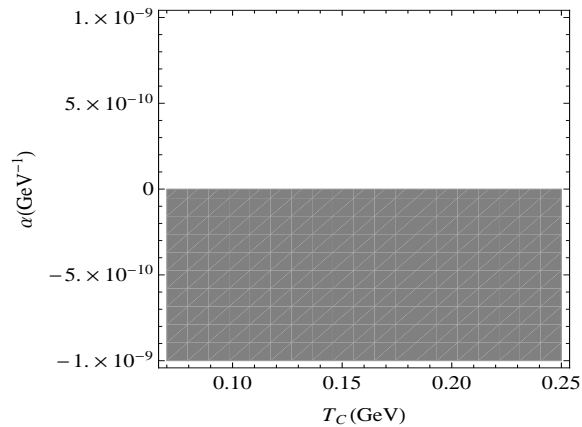,width=3in,height=2.2in,angle=0}
\end{center}
\caption{\emph{A diagram of $(T_c,\alpha)$ parameter space.}}
\label{alpha}
\end{figure}
We see that by increasing the value fixed for $T_r$, the possibility to see deviations arising from Planck scale
modification increases for $\alpha$ parameter very close to what is expected from theory.

\section{QCD sources of SGW}
\label{sec:Sources}
Since our focus in this paper is on the SGWB sourced by the first order hadronization process,
in this section we consider two significant components involved in strong QCD phase transition which have
important role in production of the SGWB. Those two components are: \emph{``bubble collisions''} which
create some shocks in the plasma medium \cite{Kosowsky:1991ua, Kosowsky:1992rz, Caprini:2007xq, Huber:2008hg}
and  \emph{``Magnetohydrodynamic turbulence''} (MHDT) which may be produced after bubble collision in the
plasma \cite{Caprini:2006jb, Caprini:2015zlo}.
\begin{figure}
\begin{tabular}{c}\hspace{-0.5cm}\epsfig{figure=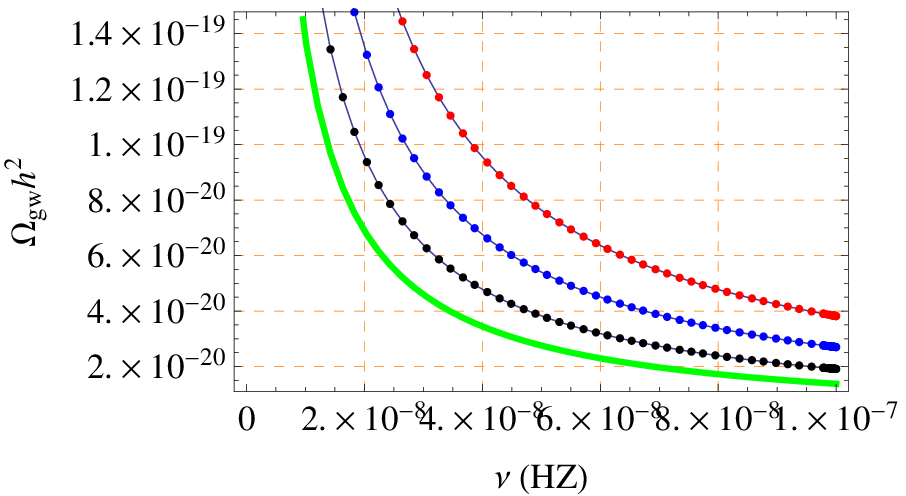, width=3.5in,height=2.2in,angle=0}
\hspace{1cm} \epsfig{figure=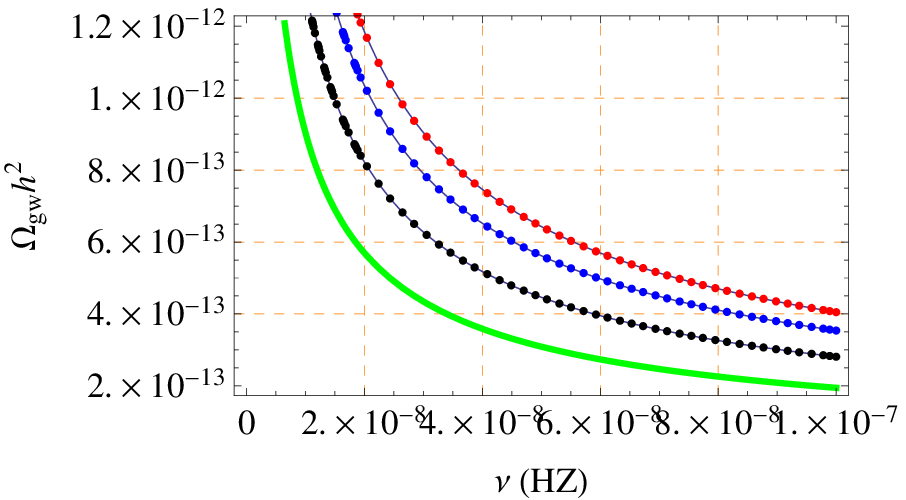, width=3.5in,height=2.2in,angle=0}
\end{tabular}
\caption{\emph{SGW signal due to the contribution of bubble collision (left panel) and
MHDT (right panel). In both of these two plots the solid line with green color refers to the case with
$\alpha=0$, while meshed lines refer to the cases with different $\alpha=-10^{-19},~-2\times10^{-19},~
-5\times10^{-19}$GeV$^{-1} $ from black to red respectively. We have set $\beta=5H_*,~u=0.7$ and also
$\frac{\kappa_b\delta}{1+\delta} = \frac{\kappa_{mhdt}\delta}{1+\delta}=0.05$ for simplicity.}}
\label{BM}
\end{figure}
In order to calculate the contribution of the SGWB spectrum produced by the bubble collisions
which the observer receives today, based on an envelope approximation the following expression has been obtained in Ref.~\cite{Huber:2008hg}
  \be \label{eq:B1}
  \Omega_{\rm gw*}^{(b)}(\nu) =\bigg(\frac{H_*}{\beta}\bigg)^2
  \bigg(\frac{ \kappa_b\delta}{1+\delta}\bigg)^2
  \bigg(\frac{0.11 u^3}{0.42 + u^2}\bigg)
 \, S_b(\nu),
  \ee
  where
  \be \label{eq:B2}
  S_b(\nu) =\frac{3.8\,(\nu/\nu_b)^{2.8}}{1 + 2.8\,(\nu/\nu_b)^{3.8}}~~~~~~\mbox{and}~~~~~~\nu_b =
  \frac{0.62 \beta}{(1.8-0.1 u + u^2)}\,\frac{a_*}{a_0} ~.
  \ee
Here, $u$ denotes the wall velocity and $\kappa_b$ is the fraction of the latent heat relevant to first order
phase transition which residues on the bubble wall. Also $\delta$ and $\beta^{-1}$ represent the ratio of the
vacuum energy density released in the phase transition relative to that of the radiation and the time duration
of the phase transition, respectively. The function $S_b(\nu)$ has been released via fitting simulation data
analytically \cite{Huber:2008hg, Jinno:2016vai} which its role is the parameterization of the the spectral shape
of the SGWB. \\

Because of high kinetic and magnetic Reynolds numbers of cosmic fluid
during hadronization process, it is expected that in the perfectly ionized plasma
medium MHDT could be produced due to percolation of the bubbles \cite{Caprini:2009yp}.
Supposing the Kolmogorov-type turbulences as raised in \cite{Kosowsky:2001xp}, the
contribution to the SGW is obtain as ~\cite{Caprini:2009yp,Binetruy:2012ze}
\be \label{eq:B3}
\Omega_{\rm gw*}^{({\rm mhdt})}(\nu) = \,\bigg(\frac{H_*}{\beta}\bigg)\,
\bigg(\frac{\kappa_{{\rm mhd}}\,\delta}{1+\delta}\bigg)^{3/2}\, u\,
S_{{\rm mhd}}(\nu),
\ee
with the parameterized function relevant to the spectral shape of the SGW spectrum as
\be
S_{{\rm mhd}}(\nu) =
\frac{(\nu/\nu_{{\rm mhd}})^3}{[(1 +\nu/\nu_{{\rm mhd}})]^{11/3}\, (1 + 8\pi\,\nu/{\cal H}_*)}\, .
\ee
Here, ${\cal H}_* = (a_*/a_0) H_*$ and $\nu_{{\rm mhd}} =\frac{7\beta}{4u} \frac{a_*}{a_0}$,
which is the peak frequency red shifted to today observer. Also $\kappa_{{\rm mhdt}}$
denotes the fraction of latent heat energy moved into the turbulence. \\

Between the Hubble parameter $H_*$ and the frequency of GW corresponding to the moment of phase transition
$\nu_*$ there is a direct relation so that a change in $H_*$  will affect the $\nu_*$. By taking the
Planck scale physics into account via the underlying UV natural cutoffs, the Hubble
parameter deviates from the standard form as follows \cite{ref:Ali2014, ref:Mohsen2017}
\be \label{eq:B3}
H=\sqrt{\frac{\rho}{3m_{P}^2}-\frac{4\alpha}{\sqrt{3}m_P}a^2\rho^{3/2}}\,,
\ee
where after combining energy conservation equation and GUP modified EOS (\ref{eq:Ee})
the following condition guarantees that the Hubble parameter in the time of phase transition
to be physically meaningful
\be \label{eq:B4}
 \alpha <\frac{\big(4\eta_1T_*^3+5\alpha \eta_2T_*^4\big)^{2/3}}{\sqrt{187 m_P^2\eta_1
 T_*^4 + 235\alpha m_P^2\eta_2 T_*^5}}\,.
  \ee
  \begin{figure}
\begin{tabular}{c}\hspace{-0.5cm}\epsfig{figure=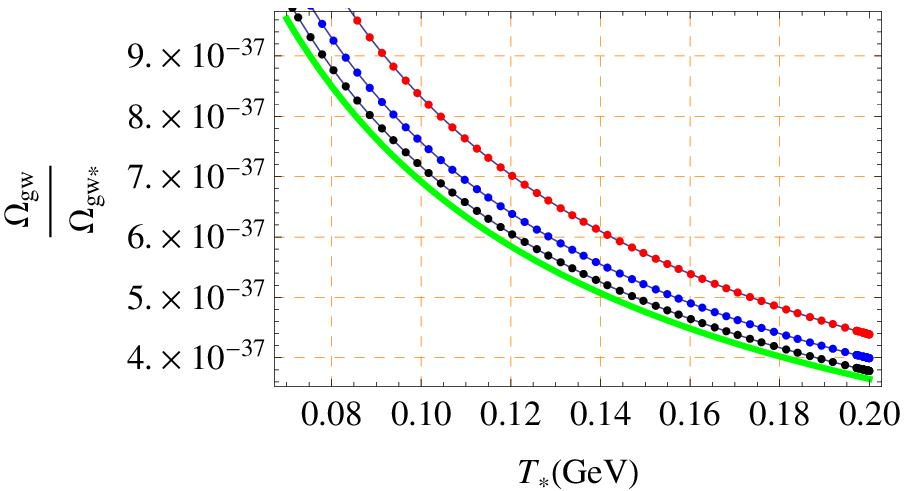, width=3.5in,height=2.2in,angle=0}
\hspace{1cm} \epsfig{figure=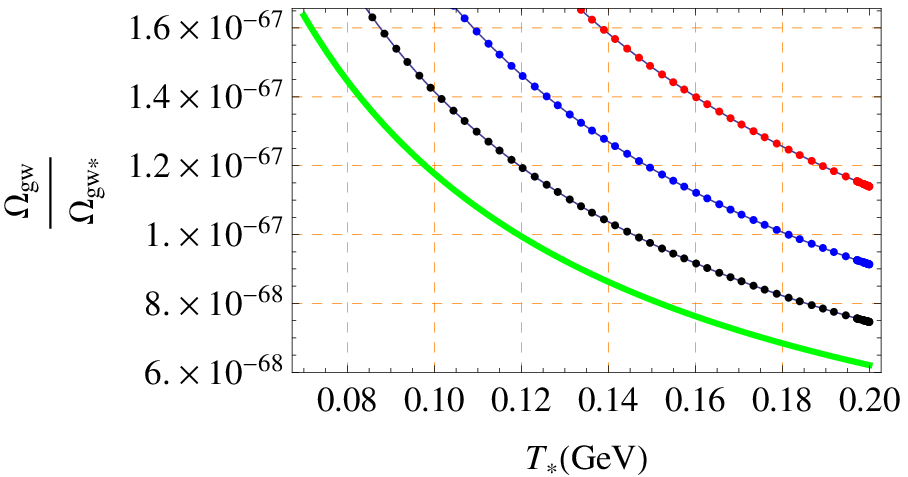, width=3.5in,height=2.2in,angle=0}
\end{tabular}
\caption{\emph{Counterpart of the middle and right panels of Fig. \ref{wz} for the case with $\alpha<0$.}}
\label{-wz}
\end{figure}
This condition leads to a $(T_c,\alpha)$ parameter space as Fig. \ref{alpha} in which just negative values of $\alpha$ are acceptable.
Despite that the negative sign of $\alpha$ seems to be unexpected at first glance, we have to encounter
with it as a suggestion in a fundamental level. It may be interesting for the reader that such a condition
also has been derived in other earlier studies, see \cite{ref:R2004, ref:P2010, ref:F2015} for instance.
In Fig. \ref{BM}, we have depicted the pure SGWB spectrum arising from bubble collision (left panel) and MDHT
(right panel) for different values of $\alpha<0$ and other involved parameters. In Fig. \ref{BM}
we have taken only the value $\beta=5H_*$. However, for larger values, the overall behavior of the plots
is the same too.  It is worth noticing that most of the QCD phase transition models suffer from a lack of accurate determination of
the duration of phase transition. In Fig. \ref{BM}, unlike Fig. \ref{wz}, an increment in the
relevant signal of GW due to incorporation of the Planck scale physics in the background evolution can be seen obviously. More technically,
the natural UV cutoffs by affecting $H_{*}$ make the transition period longer which subsequently results
in increment of the amplitude of GW signal. The reader may think that there is a contradiction between the results
obtained here and in the previous section. This is not actually the case and to achieve a unified outcome
we should recall the fact that in the previous section (see Fig. \ref{wz}) we have set positive
values for Planck scale characteristic parameter $\alpha$, while here constraint on Fig. \ref{alpha}
imposes us to consider just the negative values of this parameter. This means that the outcome of Fig. \ref{wz} in the case of setting $\alpha<0$
generally is in agreement with Fig. \ref{BM}, (see Fig. \ref{-wz}), of course with no change in the behavior
of red shift).

\section{Summary and conclusions}
\label{sec:Result}
There are different possible cosmological sources such as solitons, cosmic strings and  phase transitions,
 involved in the SGWB spectrum,  that could be potentially detected by
the present interferometric experiments. We have focused on the SGWB arising from the first
order QCD phase transition in a framework where QG effects are present. In particular,
we have used a version of GUP with the minimal length and maximum
momentum cutoffs marked by a characteristic parameter $\alpha$ on extracting the thermodynamics of ideal
quark-gluon plasma (QGP) including two massless quark flavors at  equilibrium in hadronization process
without chemical potential. Using MIT bag equation of state, modified by GUP, we have followed the calculation
of the SGWB produced in the period of QCD transition to detect possible Planck scale contributions.
We found that, due to the lack of change in the critical temperature for transition via
the GUP contribution, in the redshift of the  modified SGWB spectrum at QCD era, deviations are not observed
(see left panel of Fig. \ref{wz}). However, our calculations represent a tangible change as a drop in the
fractional energy density of the SGWB due  the UV cutoffs into MIT bag EOS, (see
middle and right panels in Fig. \ref{wz}). Indeed, here the energy density at period of transition
is an important  quantity.

Finally, we have considered two significant process: ``bubble collisions'' and
``Magnetohydrodynamic turbulence '' (MDHT),  involved in the first order QCD transition,  which contribute
in SGWB. Here, the Hubble parameter, during transition, plays a key role,  having a  physical meaning due
to GUP modifications: it  imposes  the condition $\alpha<0$ (see Fig. \ref{alpha}).
It is possible to  notice an increase in the amplitude of GWSB  without
relevant changes in redshift (see Fig. \ref{BM}). This paradox can be solved by assuming the condition
 $\alpha<0$ at fundamental level since, by replacing it without changing the redshift plot, the middle
and right panels in Fig. \ref{wz} convert to Fig. \ref{-wz} which generally is in agreement with Fig. \ref{BM}.

To conclude,  we have revisited the SGWB power spectrum, generated via first order QCD phase transition
in presence of GUP-characteristic parameter $\alpha$ with negative signature. We have derived  it without changing
the transition temperature and subsequently the redshift of QCD era, by slowing down the background evolution which  leads to an
increase in amplitude of SGWB.  This means that the probability of  SGWB signal detection, due to QCD phase transition,
could  becomes significant  for future interferometric experiments. These results  have also the potentiality to open up a new window for decrypting the dynamics of QCD transition via gravitational radiation physics.

\section*{Acknowledgments}
The work of K.N. has been financially supported by Research Institute for
Astronomy and Astrophysics of Maragha (RIAAM) under research project No. 1/5440-**.
S.C.  acknowledges the support of INFN (iniziative specifiche QGSKY and TEONGRAV) and
 COST action CA15117  (CANTATA), supported by COST (European Cooperation in Science and
 Technology).

\end{document}